\documentstyle[aps,prl,multicol,epsf,floats,epsfig]{revtex}

\begin{document}

\draft

\title{ Bose-Einstein condensation in inhomogeneous Josephson arrays}

\author{R. Burioni $^{1,2}$, D. Cassi $^{1,2}$, I. Meccoli $^{1,2}$, 
M. Rasetti $^{1,4}$, S. Regina $^{1,2}$, P. Sodano $^{1,3}$, A. 
Vezzani $^{1,2}$}

\address{ $^1$ Istituto Nazionale Fisica della Materia (INFM)\\
$^2$ Dipartimento di Fisica, Universit\`a di Parma, Italy\\
$^3$ Dipartimento di Fisica, Universit\`a di Perugia, Italy\\
$^4$ Dipartimento di Fisica, Politecnico di Torino, Italy}

\maketitle

\begin{abstract}

{We show that spatial Bose-Einstein condensation of non-interacting bosons occurs  
in dimension $d < 2$ over discrete structures with inhomogeneous topology and with 
no need of external confining potentials. Josephson junction arrays provide a physical
realization of this mechanism. The topological origin of the phenomenon may open the 
way to the engineering of quantum devices based on Bose-Einstein condensation.  
The comb array, which embodies all the relevant features of this effect, is 
studied in detail. }

\end{abstract}

\date{today}


\maketitle

\begin{multicols}{2}

\narrowtext

%%%%%%%

The recent impressive experimental demonstration of Bose-Einstein Condensation (BEC) 
\cite{STR} has stimulated a new wealth of theoretical work aimed to better understanding 
its basic mechanisms \cite{BEC} and, possibly, to exploit its consequences for
the engineering of quantum devices. 

It is well known \cite{huang} that for an ideal gas of Bose particles
BEC does not occur in dimension $d\leq 2$, and  
an $''$ad hoc$\, ''$ external confining potential is needed to reach the required density of states. 
The same is true for free bosons living on regular periodic lattices, while the result
cannot be extended to more general discrete structures lacking translational invariance.

In the following we shall prove that even for $d < 2$ \cite{note-dim} non-interacting bosons may 
lead to Bose-Einstein condensation into a single non-degenerate state, provided one resorts 
to a suitable discrete non-homogeneous support structure: indeed,  when 
the bosonic kinetic degrees of freedom do not depend on metric features only, the particles may 
feel a sort of effective interaction due to topology. 
The proposed mechanism for BEC in lower dimensional systems is then a pure effect
of the structure of the ambient space and avoids as well the need of resorting 
to external random potentials as the ones investigated by Huang in \cite{BEC}; this
is a very desirable feature in view of engineering real quantum devices.

In practice, the behavior of free bosons over generic 
discrete structures is made experimentally accessible through the realization of suitable 
arrays of Josephson junctions.  The latter are devices that can be engineered in such a way as 
to realize a variety of non-homogeneous patterns. We shall show indeed that classical 
Josephson junction arrays arranged in a non-homogeneous geometry - not even necessarily planar - 
provide an example of the proposed mechanism for BEC, leading to a single state spatial 
condensation. 

Theoretical studies of Josephson junction arrays are based on the short-range
Bose-Hubbard model, since the phase diagram of Josephson junction arrays may
be derived \cite{FWG} from an Hamiltonian describing bosons with repulsive interactions
over a lattice.  In $d=1$ the phase diagram has been studied by analytical \cite{FM} and 
quantum Monte Carlo methods \cite{BASZ}; experimentally, Josephson junction arrays are
used to study interacting bosons in one dimension. For a generic array the corresponding Hamiltonian is given by
\begin{eqnarray}
H^{BH} = U \sum_i n_i^2 + \sum_{ij} A_{ij} \left ( V n_i n_j -
J ( a_i^{\dagger} a_j + a_j^{\dagger} a_i ) \right ) \; , 
\nonumber 
\end{eqnarray} 
where $A_{ij}$ is the adjacency matrix: $A_{ij}=1$ if the sites $i$
and $j$ are nearest neighbors and  $A_{ij}=0$ otherwise; $a_i^{\dagger}$
creates a boson at site $i$ and $n_i \equiv a_i^{\dagger} a_i$. The phase
diagram structure reflects the competition between the boson kinetic energy
(hopping, favouring boson mobility) and repulsive interaction (Coulomb, working
so as to suppress dynamics). In a realistic experimental setup \cite{vOM}, the
parameters $U$ and $V$ depend on the ratio between the intergrain capacitance
$C$ and the gate capacitance $C_0$, while the parameter $J$ describes Cooper
pair hopping. Josephson junction arrays allow for a good experimental control
of $C/C_0$ and $J$, which can also be varied over a wide range.
For $U\gg J,V$, and for bipartite arrays, Hamiltonian $H^{BH}$ maps onto the
quantum spin-$\frac{1}{2}$ XXZ model \cite{AA}. On the other hand, in the weak
coupling limit ($''$classical$\, ''$ Josephson junctions) $U,V \ll J$,
realizable 
when $C/C_0 \to 0$, the hopping term dominates the physical behaviour of the
system, 
which is then described by the tight-binding Hamiltonian 
\begin{equation}
H= -t\sum_{ij}A_{ij}a^{\dagger}_i a_j \; , 
\label{ham}
\end{equation}
where $t$ is an effective hopping
parameter which accounts for a renormalization of the Josephson coupling $J$. 
For a non translation-invariant geometry of the array the tight-binding
model, which describes $''$free$\, ''$ bosons over a regular lattice, cannot 
any longer be interpreted as representing non-interacting particles, just due 
to the ambient graph topology. We shall show the dramatic effect of topology 
already 
on the simple graph referred to as square comb \cite{mattis,weiss}.  
This provides an explicit 
and remarkable example of topology-induced mechanism leading to a spatial 
Bose-Einstein condensation in low dimension. 

The $''$square comb$\, ''$ is the graph made of $N$ $''$fingers$\, 
''$ of $N$ sites represented in Fig. 1. whose total number of sites is $N^2$. In the 
following the generic vertex $i$ is labelled with the $''$coordinate$\, ''$ indices $(x_i,y_i)
\, , \, i\in {\bf Z}_N$ (where the latter requirement is imposed to guarantee periodic boundary conditions).

%begin{figure1}

\hspace{1cm}

\begin{center}
\epsfig{file=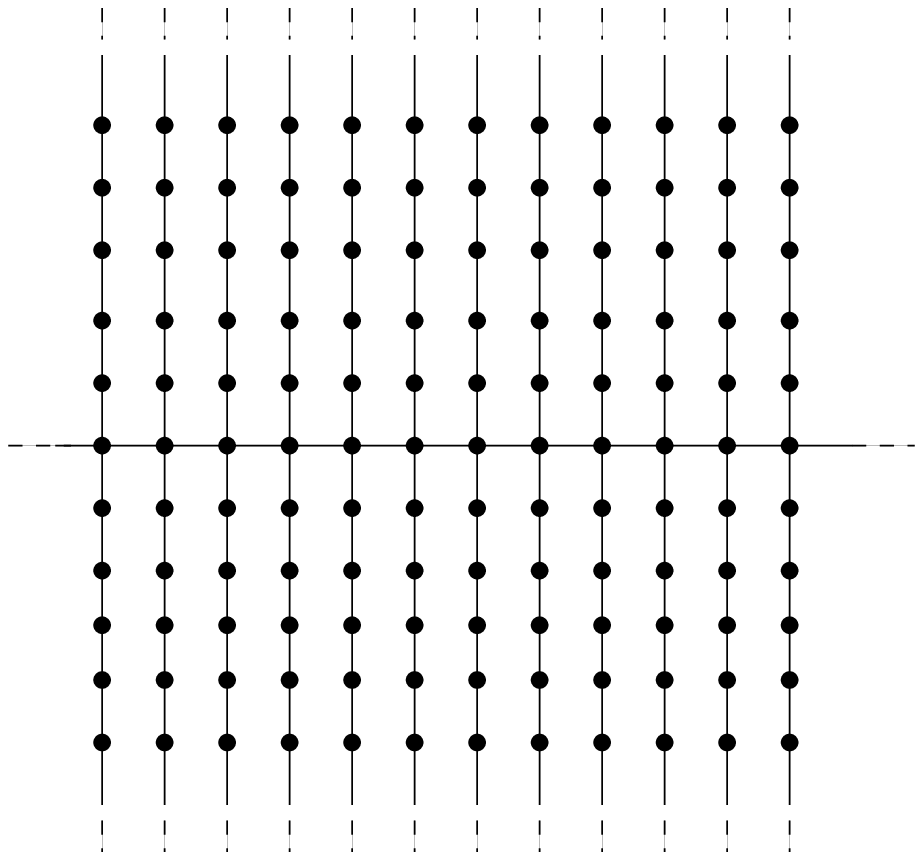,height=6cm}
\end{center}

{\noindent Fig. 1. The comb graph: the dots represent Josephson junctions and 
the links describe the topology of the array connections, with no reference to 
their embedding in real euclidean space. }

\hspace{1cm}
  
\hspace{1cm}

%end{figure1}

%%%%%%%%%%%%%%%%%%%%%%%%%%%%%%

The tight-binding model on the comb graph, is given by the Hamiltonian
(\ref{ham}) with the adjacency matrix $A_{ij}= \left ( \delta_{x_i,x_i+1} + 
\delta_{x_i,x_i-1}  \right ) \delta_{y_i,0} \delta_{y_j,0} + \left ( \delta_{y_i,y_i+1} + \delta_{y_i,y_i-1}  \right ) \delta_{x_i,x_j}$. By exploiting the comb  
translation invariance in the direction of the backbone one can perform a 
Fourier Transform along the $x$ direction obtaining a new Hamiltonian in 
the variables $k_x$:
\begin{equation}
H=-t \sum_{y,y'}\sum_{k_x}(\delta_{y,0}\delta_{y,y'}\cos k_x +{\bar A}_{y y'})
a_{k_x,y}^{\dagger} a_{k_x,y'} \; , 
\label{ham2}
\end{equation}
where $k_x=2 \pi n/N$ and $n=0,1,\cdot,N-1$ and ${\bar A}_{y y'}$ in the
adjacency matrix for a linear chain (i.e. for each comb finger). The operator $a_{k_x,y}$ is
given by: $a_{k_x,y}\equiv \sum_{x=1}^N e^{i k_x x}a_{(x,y)}$. 
Notice that (\ref{ham2}) is the sum of $N$ commuting Hamiltonians
representing an one-dimensional tight-binding model with a local potential 
at site $0$ of value: $-t \cos k_x$. 

Each of the one-dimensional Hamiltonians appearing in eq. (\ref{ham2}) can be 
diagonalized, in the thermodynamic limit.  To do this one uses   
the property that for $y\geq 2$ and $y\leq -2$ the eigenvectors of 
$-(\delta_{y,0}\delta_{y,y'}\cos k_x +{\bar A}_{y y'})$ are those of 
the Hamiltonian describing a free particle on the linear chain.
Such eigenvectors are: $e^{\pm i k_y y}$ with eigenvalue $-2 \cos k_y$,
$e^{\pm k_y y}$ with eigenvalue  $-2\cosh k_y$ and $(-1)^{y}e^{\pm k_y y}$ 
with eigenvalue $2\cosh k_y$. If one requires that the eigenvalue equations hold 
also at sites $-1$, $0$,$1$ and that the eigenvectors are normalizable, one finds
that the spectrum of the one dimensional problem is given by the isolated point
$E=-2t\,{\rm sgn}(\cos k_x)\sqrt{1+\cos^2 k_x}$ and by a continuous
part $R_0=\{E \, |\,|E|<2t\}$ with a density of states given by
$\rho_0 (E)= (1/\pi)(N-1)(4 t^2-E^2)^{-1/2}$.  
From the spectra of the one dimensional problems corresponding to different
values of $k_x$ one obtains in the thermodynamic limit
the density of states for the tight-binding Hamiltonian on the comb-graph 
(Fig. 2). 

%begin{figure2}

\hspace{1cm}

\begin{center}
\epsfig{file=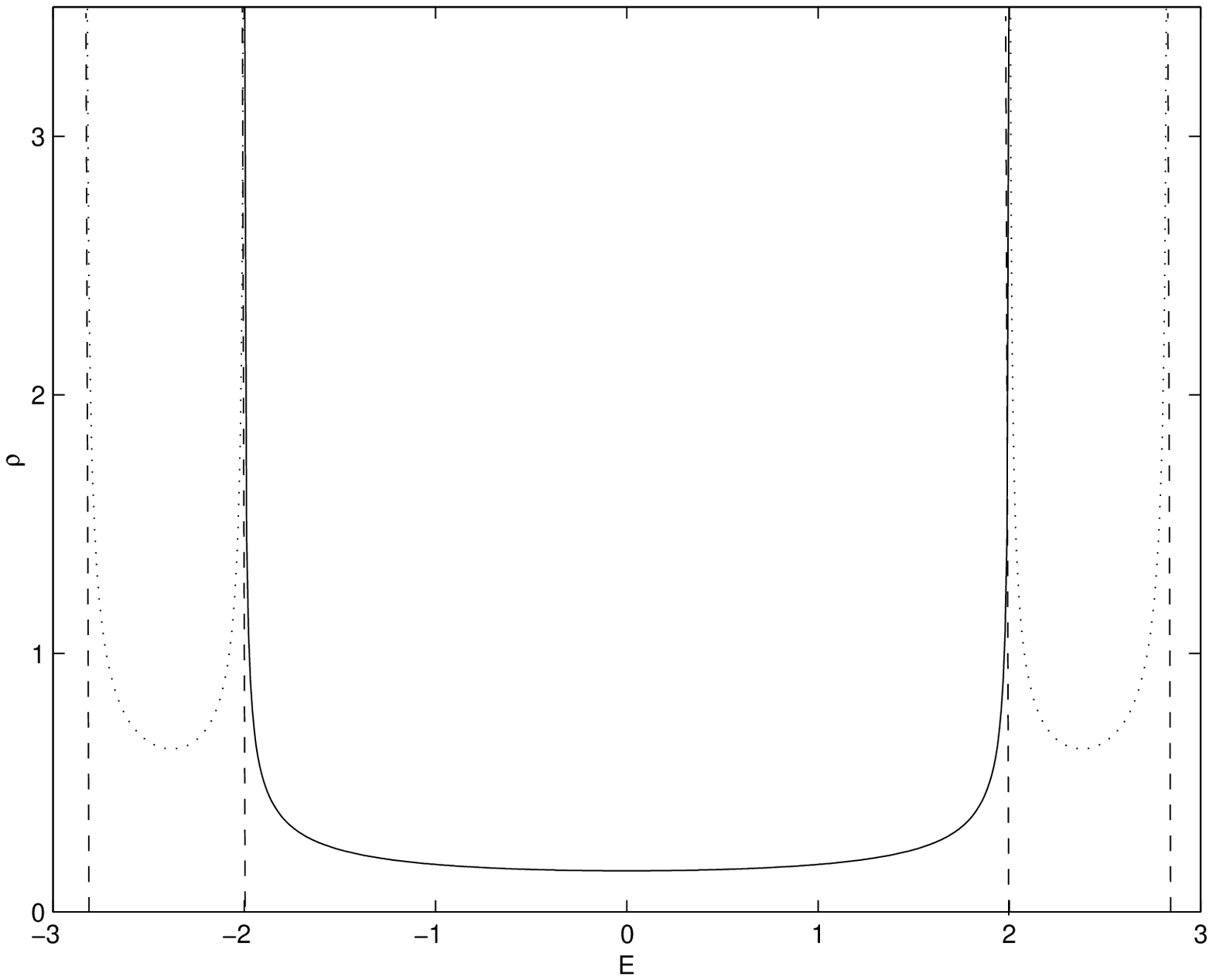,height=6cm}
\end{center}

%caption
{\noindent Fig. 2. The density of states of the Hamiltonian (\ref{ham}). The 
solid line indicates the continuous part of the spectrum $\rho_0$, which is 
normalized to $N(N-1)$. The dot lines denote the sets of zero 
measure, the densities $\rho_+$ and $\rho_-$ are normalized to $N$. The $x$-axis scale is in
units of $t$.}

\hspace{1cm}

\hspace{1cm}

%end{figure2}

%%%%%%%%%%%

The spectrum is made of three parts $R_\alpha \, , \, \alpha = 0,\pm $ : for 
$E\in R_0$ the density of states is given by:
\begin{eqnarray}
\rho_0(E)= \frac{1}{\pi}N(N-1)({4t^2-E^2})^{-1/2} \; .
\nonumber
\end{eqnarray}
Since $\int_{R_0} \rho_0(E) dE = N(N-1)$, in the thermodynamic limit, almost
all the states, i.e. all the states apart a set of measure zero, belong to
this region and $\lim_{N\to \infty} (\int_{R_0} \rho_0(E) dE)/N^2=1$.
In the other two regions, $R_-=\{E |-\sqrt{8} t\leq E <-2t\}$ and $R_+=\{E
|2t< E\leq \sqrt{8}t\}$, the density of states is given by:
\begin{eqnarray}
\rho_-(E)=\rho_+(E)\equiv\rho_\pm(E)= hN \frac{|E|}{\sqrt{8t^2-E^2}
\sqrt{E^2-4t^2}} \; , 
\nonumber
\end{eqnarray}
where $h$ must be chosen so that $\int_{R_-\cup R_+} \rho_\pm(E) dE 
= N$. There follows that in the thermodynamic limit only a subset of states
of measure zero belongs to these regions of the spectrum, in that one has
$\lim_{N \to \infty} N^{-2} \int_{R_\pm}\rho_\pm(E) dE=0$.  The states with
$E\in R_-$ play a fundamental role in the study of bosonic particles on comb
structures. The lowest energy eigenstate of (\ref{ham}) (corresponding to
$E_0=-\sqrt{8}t$) is represented in Fig. 3. It is constant along the $x$
direction for any fixed $y$, while along $y$ it decreases exponentially
with the distance from the backbone.

%begin{figure3}

\hspace{1cm}

\begin{center}
\epsfig{file=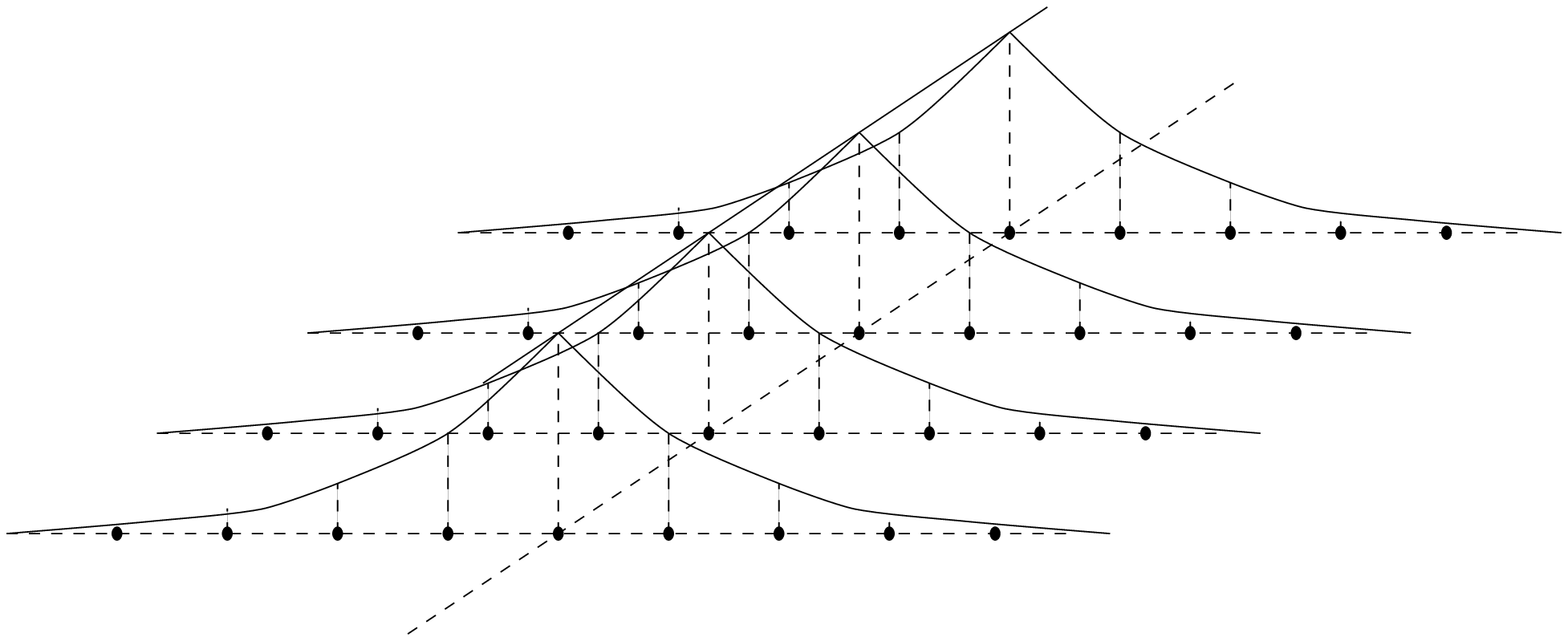,height=8cm,angle=-90}
\end{center}

%caption
{\noindent Fig. 3. The eigenvector corresponding to the lowest energy state. 
It is constant along the backbone direction and it decreases exponentially 
as $\exp(-{\rm arcsh}(2)\cdot |x|)$ along the fingers.}

\hspace{1cm}

\hspace{1cm}

%end{figure3}

%%%%%%%%%%%%%%

If one introduces a finite bosonic filling $f$, i.e. if one fills the comb with
$fN^2$ non interacting bosons, fixing the number of particles in the gran canonical
partition function amounts to choosing the fugacity $z(N,\beta,f)$ as 
\begin{equation}
f=N^{-2} \sum_{\alpha=-,0,+} \int_{R_\alpha} {1 \over z^{-1}e^{\beta E} -1} 
\rho_\alpha(E) dE \; , \label{z}
\end{equation}
with $0<z\leq e^{-\beta t \sqrt{8}}$. Since 
$\int_{R_+} (z^{-1}e^{\beta E} -1)^{-1} \rho_+(E) dE < c N$, 
($c$ is a number independent of $N$ and $z$), the third 
term of (\ref{z}) vanishes in the thermodynamic limit. On the other hand, 
the first term can be positive and finite in the thermodynamic limit if
$z(N,\beta,f) \rightarrow  e^{-\beta t \sqrt{8}}$ when $N\rightarrow \infty$. 
Denoting by $n_0$ the fraction of particles with energy smaller 
than $-2t$,  $n_0=\lim_{N\to \infty} f^{-1}N^{-2}\int_{R-} (z^{-1}(N)e^{\beta E} 
-1)^{-1} \rho_-(E) dE$, one has that $n_0$ is a finite fraction of particle condensed 
in a subset of states of measure zero.  In the thermodynamic limit, if $\beta_c$ is 
the inverse temperature for which:
\begin{eqnarray}
1={1\over \pi f}
\int_{R_0} {dE \over (e^{\beta_c (E+t\sqrt{8})} -1) 
\sqrt{4t^2-E^2}} \; , 
\nonumber
\end{eqnarray}
one has that if $\beta\leq \beta_c$ there always exixts a real and positive $z$ 
solution of the equation
\begin{eqnarray}
f= {1\over\pi}\int_{R_0} {dE \over (z^{-1}e^{\beta E} -1) \sqrt{4t^2-E^2}} \; , 
\nonumber
\end{eqnarray}
and $n_0=0$: there is no condensation. For $\beta > \beta_c$ one has
$z=e^{-\beta t \sqrt{8}}$ and $n_0$ is given by
\begin{eqnarray}
n_0=1-{1\over f\pi}
\int_{R_0}{dE \over (e^{\beta (E+t\sqrt{8})} -1) \sqrt{4t^2-E^2}} \; . 
\nonumber
\end{eqnarray}
For  $T<T_c\equiv(\beta_c)^{-1}$ there is Bose-Einstein condensation and in 
Fig. 4 we plot the fraction $n_0$ of particles in the condensate  as a 
function of $T$ for several values of the filling ($f=0.5,1,2$).
The points where the curves intersect the T-axis are 
the critical temperatures for the different fillings.

%begin{figure4}

\hspace{1cm}

\begin{center}
\epsfig{file=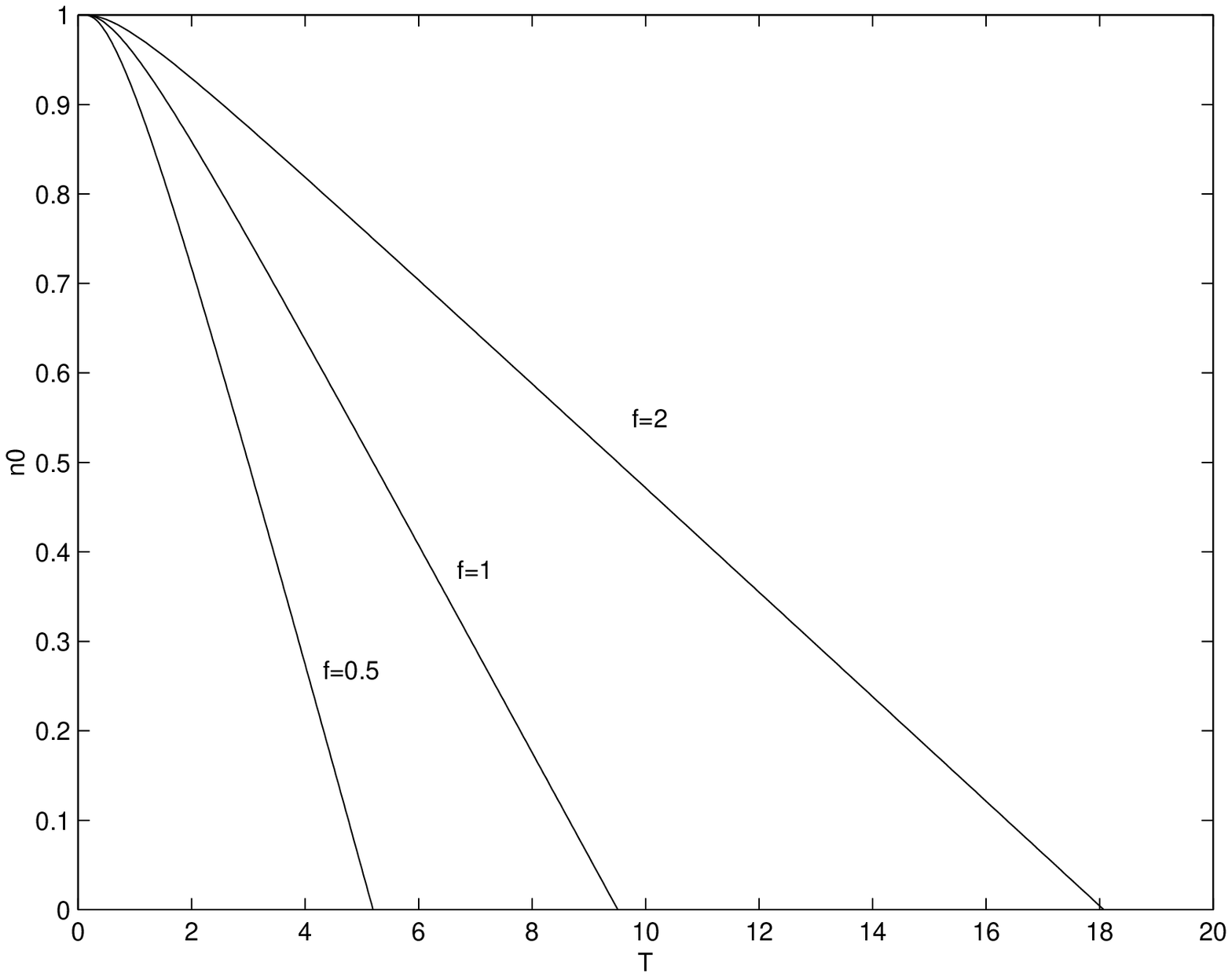,height=6cm}
\end{center}

%caption
{\noindent Fig. 4. The fraction $n_0$ of condensed particles as a function of the 
temperature $T$ for different fillings $f=0.5,1,2$.}

\hspace{1cm}

\hspace{1cm}

%end{figure4}
%%%%%%%%%%%%%%%%%

For $T\to T_c-$ the order parameter depends linearly on $T$ since $n_0\propto 
(T-T_c)$, and it is analytic in $T$ for $T$ close to $T_c$. The behaviour for 
$T\to 0$ is given by $n_0 \propto \sqrt{T}\exp(-(2\sqrt{2}-2)t/KT)$. This 
is different from the customary Bose-Einstein condensation in a 3-$d$ box, 
where $n_0=1-(T/T_c)^{3/2}$. The critical temperature $T_c$ exhibits the dependence 
on the filling $f$ shown in Fig. 5, asymptotically ($f > 1$) linear and of the form 
$T_c \propto - \bigl ( \ln f \bigr )^{-1}$ for $f \ll 1$. Both behaviours characterize 
a gapped system.  The gap $(2\sqrt{2}-2)t$ measures the difference 
between the ground state energy and the bottom of the spectral region $R_0$

%begin{figure}

\hspace{1cm}

\begin{center}
\epsfig{file=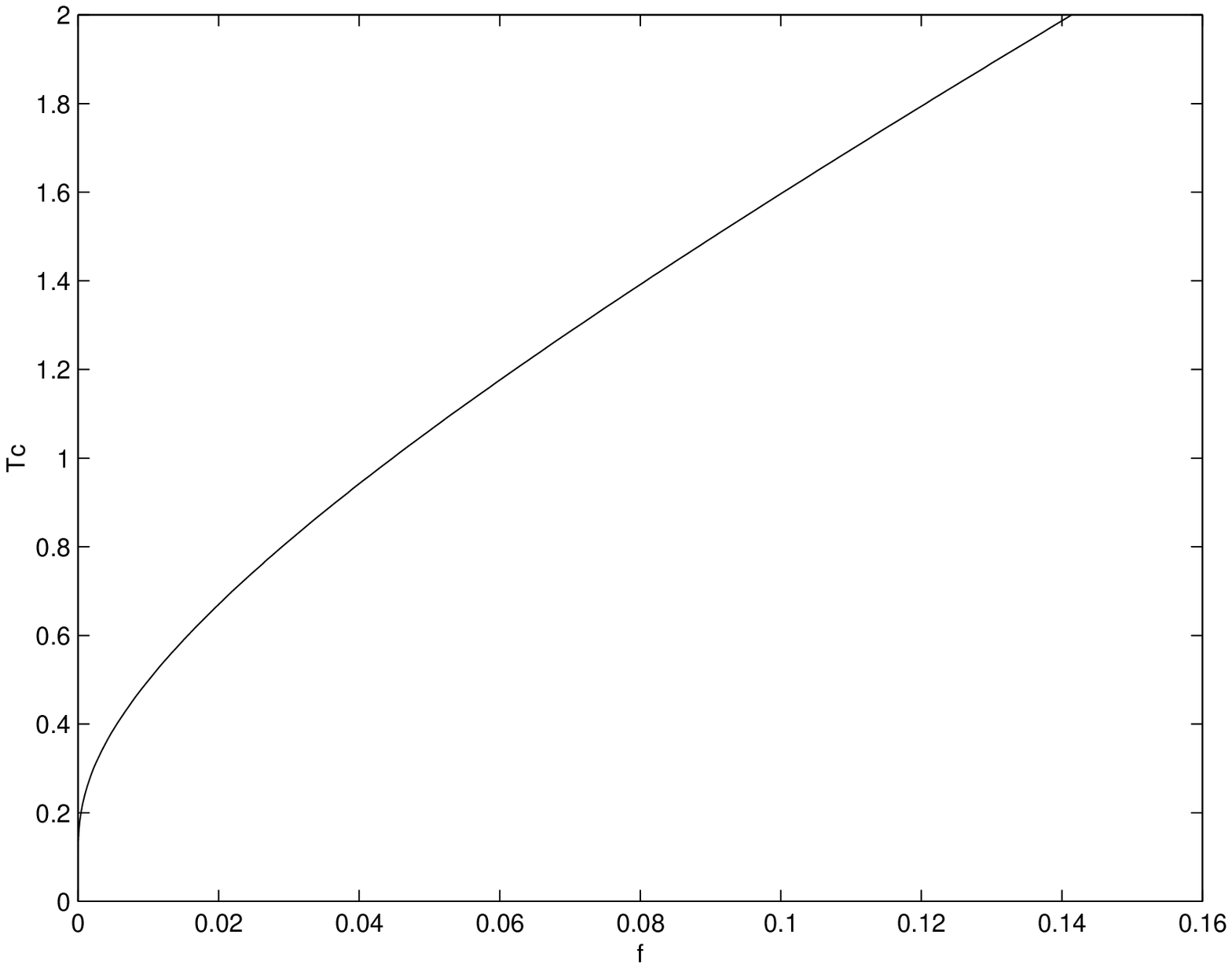,height=6cm}
\end{center}
%caption
{Fig. 5. The critical temperature $T_c$ as a function of the 
filling $f$.}

\hspace{1cm}

\hspace{1cm}

%end{figure}

%%%%%%%%%%%%%%%%%%%

The average energy per particle
\begin{eqnarray}
<E>=-\sqrt{8} t n_0+{1 \over f\pi} 
\int_{R_0}{ E dE \over (z^{-1}e^{\beta E} -1) \sqrt{4t^2-E^2}}
\; , 
\nonumber
\end{eqnarray}
shows that all the particles in the condensate have energy
$-t\sqrt{8}$, namely they are all in the ground state. 
Summarizing, in the thermodynamic limit and for $T>T_c$ 
almost all particles have energies between $-2t$ and $2t$ with the
distribution  $(z^{-1}e^{\beta E} -1)^{-1}\rho_0(E)$, while 
for $T<T_c$ a finite fraction $n_0$ of particles  
is condensed in the state of lowest energy $E_0=-\sqrt{8}t$. These
particles occupy the inhomogeneous state described in Fig. 3, in which  
the sites closer to the backbone have larger filling than the farther ones.

In conclusion, we  exhibited an explicit example (comb array)  of Bose-Einstein 
condensation into a single state for non-interacting bosons, induced in 
dimension $d< 2$ by inhomogeneities without disorder and with no confining
external potential in the discrete 
geometrical structure.  The ensuing condensate shows deconfinement in one 
direction, i.e. along the comb backbone, and localization along the orthogonal 
direction; this is expected to lead to detectable singularities in the response functions. 

The model Hamiltonian used  is physically  implementable 
by classical Josephson junction arrays, which it is possible to engineer  
in any desired geometric setting. Thus Bose-Einstein condensates arise
as an intrinsic device feature, without need of fine-tuning any external 
control parameter. 

Furthermore, it has been evidenced that a comb graph may emerge as a relevant
geometrical structure if one consider a chain of classical Josephson junctions
aligned along the direction of the backbone and interacting with a suitable
external environment \cite{schmid}. 

Finally, the same devices, as recently proposed \cite{JABR},  might lend themselves to be 
used for the realization of BEC-based encoding and manipulation 
of quantum information\cite{YuShi}. 
The idea here is that because of the spontaneous symmetry breaking
that characterizes it, a Bose-Einstein condensate should be quite naturally
described by a non-linear quantum mechanics. Such non-linearity can be thought
of as due just to the effective interaction of the bosons composing the
condensate. In such a case the scenario recently described by Abrams and Lloyd\cite{ABLO} 
whereby non-linear quantum mechanics in the sense of Weinberg\cite{WEI} implies polynomial 
time efficiency in dealing with {\bf NP}-complete and $\sharp${\bf P} complex
computations would hold. It is intriguing that for the system presented here
the relevant non-linearity, indeed present, is due to geometry and topology
rather than to physical interactions.

\end{multicols} 
\end{document}